\documentclass[english, aps, twocolumn, floatfix]{revtex4-2}
\usepackage[T1]{fontenc}
\usepackage[utf8]{inputenc}
\usepackage{color}
\usepackage{babel}
\usepackage{nicefrac}
\usepackage{amsmath}
\usepackage{amssymb}
\usepackage{graphicx}
\usepackage{microtype}
\usepackage{pgfkeys}
\usepackage{acronym}
\usepackage[unicode=true,
 bookmarks=true,bookmarksnumbered=false,bookmarksopen=false,
 breaklinks=false,pdfborder={0 0 0},pdfborderstyle={},backref=false,colorlinks=true]
 {hyperref}
\hypersetup{
    pdftitle={Two-way photoeffect-like occupancy dynamics in a single (InGa)As quantum dot},
    pdfauthor={Pavel Sterin, Kai H{\"u}hn, Jens H{\"u}bner, Michael Oestreich}
}   

\usepackage{xcolor}
\definecolor{blue}{rgb}{0.1216, 0.4667, 0.7059}
\definecolor{orange}{rgb}{1.0000, 0.4980, 0.0549}
\definecolor{green}{rgb}{0.1725, 0.6275, 0.1725}
\definecolor{red}{rgb}{0.8392, 0.1529, 0.1569}
\definecolor{violet}{rgb}{0.5804, 0.4039, 0.7412}
\definecolor{brown}{rgb}{0.5490, 0.3373, 0.2941}
\definecolor{pink}{rgb}{0.8902, 0.4667, 0.7608}
\definecolor{gray}{rgb}{0.4980, 0.4980, 0.4980}
\definecolor{olive}{rgb}{0.7373, 0.7412, 0.1333}
\definecolor{cyan}{rgb}{0.0902, 0.7451, 0.8118}

\makeatletter

\def\doibase{\doibasefix}
\def\doibasefix#110.{https://doi.org/10.}

\usepackage{siunitx}[onecolumn]

\newacro{QD}{quantum dot}
\newacro{PSD}{power spectral density}
\newacro{SNS}{spin noise spectroscopy}
\newacro{SCTS}{separation of correlator time scales}

\newcommand{\figref}[1]{Fig.~\ref{#1}}
\newcommand{\secref}[1]{Sec.~\ref{#1}}
\renewcommand{\eqref}[1]{Eq.~(\ref{#1})}

\newcommand{\citer}[1]{Ref.~\cite{#1}}
\newcommand{\citerp}[1]{Refs.~\citep{#1}}

\makeatother

\newcommand{\gammaA}{\ensuremath{\gamma_{\mathrm{a}}}}
\newcommand{\gammaR}{\ensuremath{\gamma_{\mathrm{r}}}}
\newcommand{\gammaX}{\ensuremath{\gamma_{\mathrm{p}}}}
\newcommand{\gammaH}{\ensuremath{\gamma_{\mathrm{h}}}}
\newcommand{\gammaE}{\ensuremath{\gamma_{\mathrm{e}}}}

\newcommand{\gammaC}{\ensuremath{\gamma_{\mathrm{c}}}}

\newcommand{\gammaS}{\ensuremath{\gamma_{\mathrm{s}}}}
\newcommand{\gammaN}{\ensuremath{\gamma_{\mathrm{n}}}}
\newcommand{\lambdaS}{\ensuremath{\lambda_{\mathrm{s}}}}
\newcommand{\lambdaN}{\ensuremath{\lambda_{\mathrm{n}}}}

\newcommand{\gammaD}{\ensuremath{\gamma_\mathrm{d}}}
\newcommand{\gammaTr}{\ensuremath{\gamma_{0}}}
\newcommand{\gammaSat}{\ensuremath{\gamma_{1}}}
\newcommand{\gammaNp}{\ensuremath{\gamma_{\mathrm{n}_1}}}
\newcommand{\gammaEff}{\ensuremath{\gamma_\mathrm{eff.}}}

\newcommand{\opH}{\ensuremath{\opN _\mathrm{h}}}
\newcommand{\opHD}{\ensuremath{\opN _{\nicefrac{-3}{2}}}}
\newcommand{\opHU}{\ensuremath{\opN _{\nicefrac{+3}{2}}}}
\newcommand{\opHUD}{\ensuremath{\opN _{\nicefrac{\pm 3}{2}}}}
\newcommand{\opE}{\ensuremath{\opN _\mathrm{e}}}
\newcommand{\opED}{\ensuremath{\opN _{\nicefrac{-1}{2}}}}
\newcommand{\opEU}{\ensuremath{\opN _{\nicefrac{+1}{2}}}}
\newcommand{\opEUD}{\ensuremath{\opN _{\nicefrac{\pm 1}{2}}}}
\newcommand{\opSigmaH}{\ensuremath{\hat{\sigma}_\mathrm{h}}}
\newcommand{\opSigmaE}{\ensuremath{\hat{\sigma}_\mathrm{e}}}
\newcommand{\opRot}{\ensuremath{\hat{\theta}_{K}}}
\newcommand{\opS}{\ensuremath{\hat{S}_{z}}}
\newcommand{\opN}{\ensuremath{\hat{n}}}

\newcommand{\coefCN}{\ensuremath{C_\mathrm{n}^\mathrm{i}}}
\newcommand{\coefCS}{\ensuremath{C_\mathrm{s}^\mathrm{i}}}

\begin{document}
\title{Two-way photoeffect-like occupancy dynamics in a single (InGa)As quantum dot}
\author{Pavel Sterin}
\author{Kai H{\"u}hn}
\author{Jens H{\"u}bner}
\email{jhuebner@nano.uni-hannover.de}
\author{Michael Oestreich}
\affiliation{Institut f{\"u}r Festk{\"o}rperphysik, Leibniz Universit{\"a}t Hannover, Appelstraße 2, 30167 Hannover, Germany}
\affiliation{Laboratory of Nano and Quantum Engineering, Leibniz Universit{\"a}t Hannover, Schneiderberg 39, 30167 Hannover, Germany}
\date{\today}
\begin{abstract}
    We extend optical spin noise spectroscopy on single (InGa)As quantum dots to high magnetic fields at which the splitting between the two optical active Zeeman branches of the positively charged quantum dot trion transition is significantly larger than the homogeneous line width.
    Under such conditions, the typical theoretical approximations concerning the decoupling of spin and charge dynamics are in general not valid anymore and the Kerr fluctuations show  significantly richer detuning-dependent features in the spectral region between the two Zeeman branches.
    A comparison of the experimental data with an extended theory suggests that the typical Auger-recombination can be neglected at high magnetic fields in favour of a probe-laser induced photoeffect that shuffles not only the resident hole out of the quantum dot but also activates acceptor-bound holes which recharge the empty quantum dot.
\end{abstract}
\maketitle

\section{Introduction}
Single semiconductor quantum dots (QDs), which are charged by a single electron or hole, are promising candidates for high-quality solid-state spin-photon interfaces \citep{warburton_single_2013}.
These QDs behave like artificial atoms whose properties and degrees of freedom can be preset during growth and later manipulated using quantum optical methods \citep{gerardot_optical_2008,bodey_optical_2019,giesz_coherent_2016,podemski_probing_2016}.
The most prominent and best-studied material class of such semiconductor QDs are molecular beam epitaxy grown, self-organized (InGa)As QDs, which are often naturally charged with a hole due to unavoidable $p$-type background doping by carbon impurities.
The spin of such a localized hole can reach remarkably long spin relaxation times due to the $p$-type wavefunction of the heavy hole state \citep{prechtel_decoupling_2016,de_greve_ultrafast_2011} and is thereby an especially interesting ingredient in view of optically active spin devices.

This kind of spin-photon interfaces are in general driven in a quasi-resonant excitation regime at which the laser-photon energy drives the QD transition but is below the bandgap of the QD barrier.
The quasi-resonant excitation avoids band-to-band excitation of free electrons and holes in the barrier which is known to alter the spin dynamics of the localized carrier and the optical properties of the QD \citep{moskalenko_effective_2008,holtz_optical_2012}.
However, any realistic QD is subject to additional optically induced charge transfer from and into the QD.
These additional charge transfer mechanisms are based on the Auger recombination and the photoeffect and increase the intrinsic spin relaxation rate of the localized carrier in the QD because these processes are usually incoherent \citep{lochner_internal_2021,lochner_real-time_2020,efros_random_1997,kurzmann_optical_2019}.

Spin noise spectroscopy (also known as Kerr rotation noise spectroscopy) is known to be a very efficient method for studying not only the spin, but also the occupancy dynamics in single semiconductor QDs in the regime of negligible and finite quasi-resonant optical excitation \citep{dahbashi_measurement_2012,dahbashi_optical_2014,wiegand_spin_2018,wiegand_hole-capture_2018,smirnov_nonequilibrium_2017,smirnov_theory_2021}.
So far, \ac{SNS} on single QDs has been limited to low magnetic fields and included only the Auger process as an optically induced charge transfer channel of the localized hole out of the QD.
In this paper, we extend \ac{SNS} on single QDs to high magnetic fields and include the photoeffect in two ways.
Firstly, the laser light shuffles the localized hole out of the positively charged QD and, secondly, the photoeffect delocalizes holes from unintentional background acceptors adjacent to the QD and thereby recharges the QD.
The extension to high magnetic fields is interesting from the theoretical point of view since the typical approximation of the regression analysis of separable spin and charge noise partially breaks down.
We show that this break down of the separation of correlator timescales has a significant influence on the evaluation of the experimental data by which the informative values of the single \ac{SNS} spectra degrade.
The extension to high magnetic fields is also interesting since the Auger effect is suspected to be magnetic field dependent and decreases with increasing magnetic field \citep{mannel_auger_2021}.
A distinction between Auger effect and photoeffect is possible in \ac{SNS} since the respective detuning dependences with regard to the positively charged QD trion ($X^+$) transition are different, i.e., the Auger effect depends on the effective optical excitation of the QD and thereby strongly on the detuning, while the photoeffect is detuning independent over a wide range of energies.
In fact, detailed \ac{SNS} measurements with an improved signal-to-noise ratio of about one order of magnitude in comparison to previous single QD \ac{SNS} measurement show that the Auger effect can be neglected at high magnetic fields, i.e., the photoeffect becomes the most efficient process for carrier transfer at high magnetic fields.

The paper is structured as follows.
In Sec.~\ref{sec:model_and_theory}, we amend the phenomenological model of \citerp{wiegand_spin_2018, wiegand_hole-capture_2018}, incorporate the photoeffect induced occupancy dynamics, and provide an approximate and a more exact solution. Here the approximate solution yields a more intuitive physical interpretation of the experimental and theoretical data.
In Sec.~\ref{sec:regression_breakdown}, we study in which regions the approximate solution is valid and in which regions the applied regression breaks down.
In \secref{sec:parameterization}, we discuss within the approximative model the signatures of the underlying charge transfer models in the \ac{SNS} spectra and compare in \secref{sec:experiment} our experimental results with these predictions to identify the underlying physical processes of charge transfer in natural, ungated (InGa)As QDs.

\section{Model \& Theory}\label{sec:model_and_theory}
\begin{figure}[pt!]
    \includegraphics{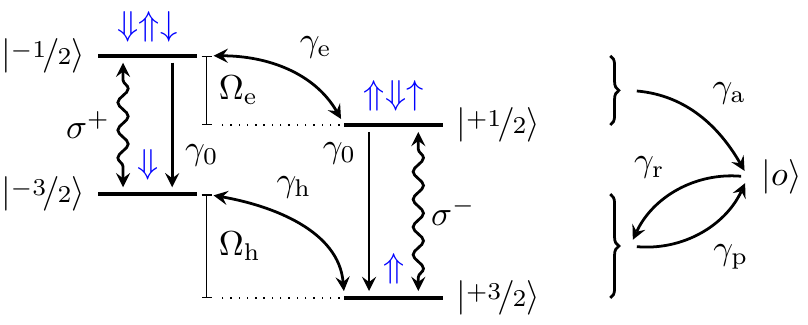}
    \caption{\label{fig:theory_model}
        Schematic of the Zeeman split \ac{QD} dynamics under resonant excitation with either $\sigma^+$ or $\sigma^-$ light.
        The blue arrows depict the electron ($\uparrow$) and hole ($\Uparrow$) spin orientations, respectively.
        The outer state is labeled by $\left|o\right>$.
        The specified parameters are the Zeeman splitting of the ground (excited) state $\Omega_\mathrm{h}$ ($\Omega_\mathrm{e}$),
        the spontaneous trion decay $\gammaTr$,
        the hole (electron) spin decay $\gammaH$ ($\gammaE$),
        the excited (ground) state loss rate $\gammaA$ ($\gammaX$),
        and the ground state reoccupation rate $\gammaR$.
    }
\end{figure}
In the following, we extend the theoretical model of \citet{wiegand_spin_2018, wiegand_hole-capture_2018} on spin and charge noise of a single hole in a QD towards a) the case of high external magnetic fields at which the Zeeman branches of the $\mathrm{X}^{+}$ transition are split by more than an optical line width and b) occupancy dynamics induced by photoeffect-like excitation of localized holes.
The left part of Fig.~\ref{fig:theory_model} depicts schematically the relevant QD states under resonant excitation \citep{smirnov_nonequilibrium_2017,wiegand_spin_2018}.
The two ground states of the four-level system are pure heavy-hole pseudo-spin $\left |\pm\nicefrac{3}{2}\right>$ states that are split at finite magnetic fields by the heavy-hole Zeeman splitting $\Omega_\mathrm{h}$.
Any influence of the light-hole spin $\pm \nicefrac{1}{2}$ states is assumed to be negligible due to the huge confinement-induced heavy-hole light-hole splitting \citep{warburton_single_2013}.
The two excited states of the four-level system are the two electron-spin-split $X^+$ ground states of the positively charged exciton complex that incorporates an electron-hole pair in addition to the resident hole.
Pauli exclusion forces the two holes in the $X^+$ ground state into a singlet with an effective spin of zero (designated by $\Uparrow\Downarrow$ for brevity in \figref{fig:theory_model}).
Therefore, the spin of the excited state is determined only by the remaining electron spin projections $\left|\pm\nicefrac{1}{2}\right>$ that are split by the electron Zeeman splitting $\Omega_\mathrm{e}$.

The interaction with light is subject to the optical selection rules.
An excited state $\left|\pm\nicefrac{1}{2}\right>$ can only be created by a $\sigma^{\mp}$ photon from a ground state $\left|\pm\nicefrac{3}{2}\right>$ of the same sign as the excited one.
We use the rotating wave approximation and utilize the rotating frame to obtain the effective \ac{QD} Hamiltonian for interaction with linearly polarized light ($\hbar = 1$) \citep{smirnov_theory_2015,smirnov_nonequilibrium_2017}:
\begin{equation}
    \mathcal{H}=-\opE\Delta +\Omega_\mathrm{h}\opSigmaH +\Omega_\mathrm{e}\opSigmaE -\frac{\mathcal{E}}{\sqrt{2}}\left(\hat{d}_{+}+\hat{d}_{-}+\textrm{H.c.}\right),
    % Hermitian conjugate -> H.c.
\end{equation}
where $\opE =\opEU +\opED $ is the total population of the excited states, $\opEU$ and $\opED$ are the respective individual populations with electron spins $\pm \nicefrac{1}{2}$, $\Delta$ is the probe laser detuning with respect to the optical resonance at zero magnetic field, $\opE\Delta$ is accordingly the atomic-like Hamiltonian, $\opSigmaH =\opHU-\opHD$ and $\opSigmaE =\opEU -\opED $ are the (pseudo) spin polarizations of the excited and ground states, $\opHU$ and $\opHD$ are the populations of the two spin split ground states, $\mathcal{E}$ is a measure of the dipole transition magnitude, and $\hat{d}_{\pm}=\left|\nicefrac{\pm3}{2}\right\rangle \left\langle \nicefrac{\pm1}{2}\right|$ are the dipole operators.
The population of the ground states, $\opH =\opHU+\opHD$, with its heavy-hole spins $\pm \nicefrac{3}{2}$ drops out in this Hamiltonian due to the usual choice of the energy origin being the ground state energy.

\subsection{Quasi-Stationary Equilibrium}
Constant quasi-resonant excitation and various relaxation processes drive the \ac{QD} system away from thermal equilibrium towards a different steady state.
Fluctuations of spin and occupancy around this new steady state can then be detected as Kerr rotation fluctuations which yield information about the spin and occupancy dynamics.
However, the general steady-state solution of this theoretical problem is rather complex.
Therefore, similar to the approach in \citer{wiegand_hole-capture_2018}, we start by separating the fast optical decay according to the time scales from all other decay processes.
This separation is possible because the other decay processes are significantly slower than the trion decay with $\gammaTr/h  \gg \SI{300}{\mega\hertz}$, which defines the time scale on which the optical signal is formed.
Consequently, we apply the density matrix formalism first to the four-level system without considering any other relaxation processes and derive a quasi-stationary solution that we then use to calculate the true steady state that includes all other decay processes.

With these assumptions, the temporal evolution of the reduced system is given by the following Liouville–von Neumann equation:
\begin{align}
    \dot{\hat{\rho}}                                    & =i\left[\hat{\rho},\mathcal{H}\right]+\mathcal{D}_{\mathrm{trion}}\left[\hat{\rho}\right]+\mathcal{D}_{\mathrm{extra}}\left[\hat{\rho}\right],
    \label{eq:DMF-evolution}
    \\
    \mathcal{D}_{\mathrm{trion}}\left[\hat{\rho}\right] & =-\gammaTr \left(\mathcal{D}_{\hat{d}_{+}}\left[\hat{\rho}\right]+\mathcal{D}_{\hat{d}_{-}}\left[\hat{\rho}\right]\right), \nonumber                          \\
    \mathcal{D}_{\mathrm{extra}}\left[\hat{\rho}\right] & =-\dfrac{\gamma}{2}\left(\mathcal{D}_{\hat{d^{\prime}}_{+}}\left[\hat{\rho}\right]+\mathcal{D}_{\hat{d^{\prime}}_{-}}\left[\hat{\rho}\right]\right),\nonumber
\end{align}
where $\hat{\rho}$ is the density operator, $\hat{d}_{\pm}^{\prime}=\opEUD -\opHUD $, $\mathcal{D}_{\mathrm{trion}}\left[\hat{\rho}\right]$ is the damping term of the trion decay and $\mathcal{D}_{\mathrm{extra}}\left[\hat{\rho}\right]$ is a damping term describing additional optical dephasing due to the other relaxation processes, where the corresponding damping superoperator for an arbitrary operator $\hat{o}$ is defined by $\mathcal{D}_{\hat{o}}\left[\hat{\rho}\right]=\frac{1}{2}\left(\hat{o}^{\dagger}\hat{o}\hat{\rho}+\hat{\rho}\hat{o}^{\dagger}\hat{o}-2\hat{o}\hat{\rho}\hat{o}^{\dagger}\right)$.

The solution of this system of differential equations can be expressed in terms of the expectation values of the populations $\opEUD $ and $\opHUD $ and the dipole operators $\hat{d}_{\pm}$:
\begin{align}
    \kappa_{\pm} & =\frac{n_{\nicefrac{\pm1}{2}}}{n_{\nicefrac{\pm3}{2}}+n_{\nicefrac{\pm1}{2}}}=\frac{\mathcal{E}^{2}\left(2\gamma+\gammaTr \right)}{2\gammaTr \left(\gammaTr ^{2}+\Delta_{\pm}^{2}\right)},
    \\
    d_{\pm}      & =\mp\dfrac{\sqrt{2}\Delta_{\pm}n_{\nicefrac{\pm3}{2}}\mathcal{E}}{\gammaTr ^{2}+\gammaD ^{2}+2\Delta_{\pm}^{2}}\pm i\dfrac{\sqrt{2}\gammaD n_{\nicefrac{\pm3}{2}}\mathcal{E}}{\gammaTr ^{2}+\gammaD ^{2}+2\Delta_{\pm}^{2}},
\end{align}
where $\gamma_0$ is the trion recombination rate, $\gamma$ is the dephasing rate due to the other relaxation processes, $\gammaD =\gamma+\frac{\gammaTr }{2}$ is the resulting effective dephasing rate, $\gammaSat =\gammaD \sqrt{1+r}$ is the width of the saturation-broadened line, $r$ is the relative intensity given by the line width and the dephasing rate ($\nicefrac{2\mathcal{E}^{2}}{\gammaD \gammaTr }$) \citep{wiegand_spin_2018,sun_non-equilibrium_2022} or by the ratio $\nicefrac{I}{I_\mathrm{sat}}$ of the probe and saturation intensities, and $\Delta_{\pm}=\Delta\mp\frac{\Omega}{2}$ with $\Omega=\Omega_\mathrm{e}-\Omega_\mathrm{h}$ is the relative detuning \citep{wiegand_hole-capture_2018}.
Transforming the dipole solution into a linearly polarized basis using $\frac{-i}{\sqrt{2}}\left(d_{-}+d_{+}\right)=d_{y}$ and expressing the result in terms of total occupancy $\opN =\opE +\opH $ and total pseudo-spin $\opS =\nicefrac{1}{2}\left(\opSigmaE +\opSigmaH \right)$ yields the expectation value of the dipole operator in the following form:
\begin{alignat}{1}
    \Im\left(d_{y}\right)= & \;\coefCS \,S_{z}+\coefCN \,n\label{eq:DMF-dyi}                                                                                                              \\
    \equiv                 & \;\dfrac{\mathcal{E}\Delta}{2}\left(\dfrac{\Delta_{-}}{\gammaTr ^{2}+\Delta_{-}^{2}}+\dfrac{\Delta_{+}}{\gammaTr ^{2}+\Delta_{+}^{2}}\right)\,S_{z}\nonumber \\
                           & \;-\dfrac{\mathcal{E}\Omega}{4}\left(\dfrac{\Delta_{-}}{\gammaTr ^{2}+\Delta_{-}^{2}}-\dfrac{\Delta_{+}}{\gammaTr ^{2}+\Delta_{+}^{2}}\right)\,n.\nonumber
\end{alignat}
This expectation value describes the refractive part of the optical response for linearly polarized light \citep{yugova_pump-probe_2009, wiegand_spin_2018} and is proportional to the Kerr rotation $\theta_K$.
The resulting Kerr rotation spectra contain the total powers of the spin and occupancy noise contributions. These contributions are in general not just proportional to the squares of the respective coefficients $\coefCS $ and $\coefCN$.
Rather, the fluctuation spectra can be expanded by mixed products of these coefficients, as we show in the following.

\subsection{True Steady State}
Next, we return to \eqref{eq:DMF-evolution} and add the damping terms that occur on a slower time scale and have been neglected so far.
We describe the hole charge dynamics using the term
\begin{align}
    \mathcal{D}_{\mathrm{charge}}\left[\hat{\rho}\right]= & -\gammaA \left(\mathcal{D}_{\hat{v}_{+}}\left[\hat{\rho}\right]+\mathcal{D}_{\hat{v}_{-}}\left[\hat{\rho}\right]\right)\nonumber                        \\
                                                          & +\gammaX \left(\mathcal{D}_{\hat{r}_{+}^{\dagger}}\left[\hat{\rho}\right]+\mathcal{D}_{\hat{r}_{-}^{\dagger}}\left[\hat{\rho}\right]\right)\nonumber    \\
                                                          & +\dfrac{\gammaR }{2}\left(\mathcal{D}_{\hat{r}_{+}}\left[\hat{\rho}\right]+\mathcal{D}_{\hat{r}_{-}}\left[\hat{\rho}\right]\right),\label{eq:DMF-auger}
\end{align}
where the rates $\gammaA$ and $\gammaX$ quantify the rate at which the hole is ejected from the excited and ground states, respectively, $\gammaR$ is the reoccupation rate of the ground states, $\hat{v}_{\pm}=\left|o\right\rangle \left\langle \nicefrac{\pm1}{2}\right|$, and $\hat{r}_{\pm}=\left|\nicefrac{\pm3}{2}\right\rangle \left\langle o\right|$.
Just as in \citer{wiegand_hole-capture_2018}, we collapse all possible states of the hole outside of the \ac{QD} into a single effective outer state $\left|o\right>$ that does not exhibit a coherent coupling to the \ac{QD} (see right part of Fig.~\ref{fig:theory_model}).
This reduction is possible, as the exact structure of the outer state is only indirectly visible in the fluctuation spectra through reoccupation with the rate $\gammaR$.
Here, the fixed rate $\gammaA$ corresponds to an Auger-like process where the exciton decays nonradiatively and the hole is moved incoherently to the outer state.
The probe intensity dependent rate $\gammaX(r)$ corresponds to an internal photoeffect-like process where a photon is absorbed and the hole is incoherently moved from the \ac{QD} to the outer state.
A fixed rate $\gammaR$ corresponds to a thermally activated process, where the hole tunnels back into the \ac{QD} ground states.
Conversely, a probe intensity dependent rate $\gammaR(r)$ corresponds to a second internal photoeffect-like process with an inverted direction where the laser field moves the hole \emph{into} the \ac{QD} ground states.
Note that we call the occupancy dynamics \emph{Auger-like} and \emph{photoeffect-like} to stress that the actual microscopic processes that invoke the loss or gain of the hole charge are not included in this phenomenological model.

We model the (pseudo-)spin relaxation in the excited and ground states using a damping term of the form
\begin{align}
    \mathcal{D}_{\mathrm{spin}}\left[\hat{\rho}\right]= & -\dfrac{\gammaE }{2}\left(\mathcal{D}_{\hat{\chi}_\mathrm{e}}\left[\hat{\rho}\right]+\mathcal{D}_{\hat{\chi}_\mathrm{e}^{\dagger}}\left[\hat{\rho}\right]\right)\nonumber \\
                                                        & -\dfrac{\gammaH }{2}\left(\mathcal{D}_{\hat{\chi}_\mathrm{h}}\left[\hat{\rho}\right]+\mathcal{D}_{\hat{\chi}_\mathrm{h}^{\dagger}}\left[\hat{\rho}\right]\right),
\end{align}
where $\hat{\chi}_\mathrm{e}=\left|\nicefrac{-1}{2}\right\rangle \left\langle \nicefrac{+1}{2}\right|$ and $\hat{\chi}_\mathrm{h}=\left|\nicefrac{-3}{2}\right\rangle \left\langle \nicefrac{+3}{2}\right|$ are the spin flip operators of the excited and ground states and $\gammaE$ and $\gammaH$ are the (pseudo-)spin relaxation rates of the electron in the excited state and the hole in the ground state, respectively.

The differential equations for the expectation values of the populations $\opN_{\pm\nicefrac{3}{2}}$, $\opN_{\pm\nicefrac{1}{2}}$, and $\opN_\mathrm{o} = \left|o\right>\left<o\right|$ can be expressed as a set of kinetic equations that govern the true steady state by including the above damping terms into \eqref{eq:DMF-evolution}:
\begin{align}
    \dot{S}_{z}  & =-\gammaS S_{z}+\lambdaN n,
    \\
    \dot{n}      & =-\gammaN n+\lambdaS S_{z}+\gammaR n_\mathrm{o},
    \\
    n_\mathrm{o} & = 1 - n.
\end{align}

Here, $n_\mathrm{o}$ is the population of the outer state and the kinetic coefficients $\gamma_\mathrm{n,s}$ and $\lambda_{s,n}$ are given by
\begin{align}
    \gammaS  & =\gammaH + \gammaX + \left(\gammaA + \gammaE - \gammaH - \gammaX\right)\frac{\kappa_{-}+\kappa_{+}}{2}
    \label{eq:gamma_s},                                                                                               \\
    \lambdaN & =\left(\gammaA + \gammaE - \gammaH - \gammaX\right)\frac{\kappa_{-}-\kappa_{+}}{4},                    \\
    \gammaN  & =\gammaX + \left(\gammaA - \gammaX\right)\frac{\kappa_{+}+\kappa_{-}}{2},                              \\
    \lambdaS & =\left(\gammaA - \gammaX\right)\left(\kappa_{-}-\kappa_{+}\right).
\end{align}
With that in mind, the steady-state expectation values for the total pseudo-spin and the \ac{QD} occupancy read as follows:
\begin{align}
    \bar{S}_{z}          & =\dfrac{\gammaR \lambdaN}{\gammaN \gammaS +\gammaR \gammaS -\lambdaN \lambdaS}, \\
    \bar{n}              & =\dfrac{\gammaR \gammaS}{\gammaN \gammaS +\gammaR \gammaS -\lambdaN \lambdaS},  \\
    \bar{n}_{\mathrm{o}} & = 1 - \bar{n} \nonumber.
\end{align}
These quantities define the steady-state density matrix that constitutes the equilibrium state upon which fluctuations arise, and all the expectation values of the various correlators below are meant to be evaluated with respect to this steady state.

The fluctuations of the Kerr signal observed in an experiment can be expressed using \eqref{eq:DMF-dyi} as:
\begin{align}
    \delta\opRot =\opRot  - \theta_K = \coefCS \,\delta\opS +\coefCN \,\delta\opN .
\end{align}
Therefore, the autocorrelation function decay spectrum of this operator is equivalent to the \ac{PSD} spectra of Kerr noise spectroscopy.
The autocorrelation function $c_{K}(\tau)$ of Kerr fluctuations
\begin{align}
    c_{K}(\tau) & = \left\langle \delta\opRot \delta\opRot (\tau)\right\rangle
    \nonumber                                                                                                                                              \\
                & = \coefCS \,\left\langle \delta\opRot \delta\opS (\tau)\right\rangle +\coefCN \,\left\langle \delta\opRot \delta\opN (\tau)\right\rangle
\end{align}
can be derived from the evolution equations of the density matrix using the quantum regression formula \citep{carmichael_open_1993,carmichael_statistical_1999}.
The autocorrelator $c_{K}(\tau)$ is governed by the kinetic equations described above because the underlying correlators are governed by the same set of differential equations as the density matrix itself.
As a consequence, the kinetic equations can be rewritten for $\delta\opRot $ (and any other system operator) as a set of differential equations for the correlators of the system operator and one of the fluctuations $\delta\opS $ and $\delta\opN $ ($\tau>0$):
\begin{align}
    \dfrac{\mathrm{d}}{\mathrm{d}\tau}\left\langle \delta\opRot \delta\opS (\tau)\right\rangle = & -\gammaS \left\langle \delta\opRot \delta\opS (\tau)\right\rangle \nonumber                       \\
                                                                                                 & +\lambdaN\left\langle \delta\opRot \delta\opN (\tau)\right\rangle
    \label{eq:DMF-dgl-cs},                                                                                                                                                                           \\
    \dfrac{\mathrm{d}}{\mathrm{d}\tau}\left\langle \delta\opRot \delta\opN (\tau)\right\rangle = & -\left(\gammaN +\gammaR \right)\left\langle \delta\opRot \delta\opN (\tau)\right\rangle \nonumber \\
                                                                                                 & +\lambdaS \left\langle \delta\opRot \delta\opS (\tau)\right\rangle
    \label{eq:DMF-dgl-cn}.
\end{align}
The initial conditions for this set of differential equations are given by the steady-state correlators as in \citer{wiegand_hole-capture_2018}:
\begin{align}
    \left\langle \delta\opRot \delta\opS \right\rangle & =\coefCN \,\bar{S}_{z}\left(1-\bar{n}\right)+\coefCS \,\left(\dfrac{\bar{n}}{4}-\bar{S}_{z}^{2}\right), \\
    \left\langle \delta\opRot \delta\opN \right\rangle & =\coefCS \,\bar{S}_{z}+\coefCN \,\bar{n}\left(1-\bar{n}\right).
\end{align}

Experimentally, it is useful to treat the correlation rate and the noise power of occupancy and spin fluctuations as independent parameters \citep{wiegand_hole-capture_2018}.
This is only possible if either the correlation rates or the noise powers of the two contributions have very different magnitudes.
We assume for the rest of this subsection that the pseudo-spin correlator decays much faster than the occupancy correlator.
As a consequence, $\gammaS $ can be interpreted as a correlation rate that dominates the decay of the autocorrelation function $c_{K}(\tau)$ for short delays.
In turn, the occupancy correlator decays slower and is adiabatically followed by the faster pseudospin correlator.
Furthermore, its own decay can only depend on the steady-state value of the faster correlator, such that
the solution of the kinetic equations can be written as
\begin{align}
    \left\langle \delta\opRot \delta\opS (\tau)\right\rangle = & \frac{\lambdaN }{\gammaS }\left\langle \delta\opRot \delta\opN (\tau)\right\rangle                                                                                          \\
                                                               & +e^{-\tau\gammaS }\left(\left\langle \delta\opRot \delta\opS \right\rangle -\frac{\lambdaN }{\gammaS }\left\langle \delta\opRot \delta\opN \right\rangle \right), \nonumber \\
    \left\langle \delta\opRot \delta\opN (\tau)\right\rangle = & e^{-\tau\gammaNp }\left\langle \delta\opRot \delta\opN \right\rangle,
\end{align}
where $\gammaNp =\gammaN +\gammaR -\frac{\lambdaS \lambdaN }{\gammaS }$ is the effective occupancy correlation rate.

\subsection{Autocorrelation Spectra}
The expected \ac{PSD} spectrum is just the Fourier transform of $c_{K}(\tau)$ which can be decomposed for the approximated case of a fast pseudo-spin correlator decay into the contributions of spin noise and occupancy noise:
\begin{align}
    \tilde{c}_{K}^{\Vert}(f) & =A_{\mathrm{SN}}\dfrac{\mathcal{E}^2\gammaS }{4f^{2}\pi^{2}+\gammaS ^{2}}+A_{\mathrm{ON}}\dfrac{\mathcal{E}^2\gammaNp }{4f^{2}\pi^{2}+\gammaNp ^{2}}\label{eq:DMF-Kerr-sol},                \\
    A_{\mathrm{SN}}          & = \frac{1}{\mathcal{E}^2} 2\left(\coefCS \left\langle \delta\opRot \delta\opS \right\rangle -\coefCS\dfrac{\lambdaN }{\gammaS }\left\langle \delta\opRot \delta\opN \right\rangle \right),  \\
    A_{\mathrm{ON}}          & = \frac{1}{\mathcal{E}^2} 2\left(\coefCN \left\langle \delta\opRot \delta\opN \right\rangle +\coefCS \dfrac{\lambdaN }{\gammaS }\left\langle \delta\opRot \delta\opN \right\rangle \right).
\end{align}
Here, $A_{\mathrm{SN}}$ and $A_{\mathrm{ON}}$ are the integrated total noise powers of the two Lorentzian-shaped contributions of spin and occupation noise with half-width-at-half-maximum values of ${\gammaS }/{2\pi}$ and ${\gammaNp }/{2\pi}$, respectively.
We call this the \ac{SCTS} solution and use the symbol $\Vert$ to designate this approximated solution.

\begin{figure}[t]
    \includegraphics{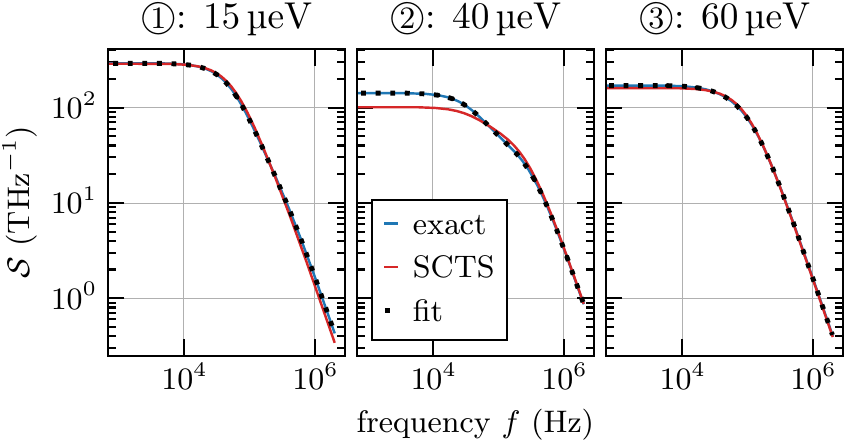}
    \caption{\label{fig:exact_failure_example}Comparison of the exact (\textcolor{blue}{blue}) and the \ac{SCTS} (\textcolor{red}{red}) solution for three different detunings $\Delta$.
        The parameters for the calculations are listed at the top of \figref{fig:exact_failure}.
        The dotted \textcolor{black}{black} curve is a nonlinear least squares fit of a double Lorentzian to the exact spectrum (the fit closely approximates the exact spectrum).}
\end{figure}

If the \ac{SCTS} is not applicable, the system of differential equations defined by \eqref{eq:DMF-dgl-cs} and \eqref{eq:DMF-dgl-cn} has to be solved exactly.
In this case, the solution can be expressed as a sum of terms $\tilde{c}_{K}(f)=\sum_{q}\mathcal{Q}_{q}$, where
\begin{align}
    \mathcal{Q}_{q} & = \frac{2q_{1}}{\omega^{2}\xi_2+\xi_1^{2}+\omega^{4}} \times \nonumber                                                          \\
                    & \left(-q_{2}q_{5}\lambdaN \lambdaS +q_{3}q_{6}\left(\xi_1-\omega^{2}\right)+q_{2}q_{4}\left(q_{5}^{2}+\omega^{2}\right)\right),
    \nonumber                                                                                                                                         \\
    q               & \in \begin{Bmatrix}
                              \begin{pmatrix}\coefCS  & \left\langle \delta\opRot \delta\opS \right\rangle  & \left\langle \delta\opRot \delta\opN \right\rangle  & \gammaS  & \gammaC  & \lambdaN \end{pmatrix} \\
                              \begin{pmatrix}\coefCN  & \left\langle \delta\opRot \delta\opN \right\rangle & \left\langle \delta\opRot \delta\opS \right\rangle  & \gammaC  & \gammaS  & \lambdaS \end{pmatrix}
                          \end{Bmatrix},\nonumber                                                                                                  \\
    \gammaC         & = \gammaN +\gammaR, \nonumber                                                                                                   \\
    \xi_1           & = \gammaC  \gammaS -\lambdaN  \lambdaS, \nonumber                                                                               \\
    \xi_2           & = \gammaC ^2+2 \lambdaN  \lambdaS +\gammaS ^2. \nonumber
\end{align}

Experimental one-sided spectra contain only positive frequencies, and hence, the powers at negative frequencies have to be accounted for by an additional factor of 2.
Furthermore, experimental noise spectra have an extrinsic quadratic optical probe power dependence \citep{hubner_rise_2014} that is usually normalized to make spectra taken at different intensities comparable.
Consequently, we divide the model spectrum by the factor $\mathcal{E}^2$ to make its area, i.e., total noise power, laser power independent in the same way.
The corresponding model quantity is then
\begin{align}
    \mathcal{S}(f) = A_{\mathrm{PSD}} \frac{2}{\mathcal{E}^2} \tilde{c}_{K}(f),
\end{align}
where $A_{\mathrm{PSD}}$ is a constant that quantifies the sensitivity of the experiment.
In the following, we employ the convention of designating experimentally accessible quantities with a star ($\star$).
The spectrum predicted by the exact solution should correspond directly to its experimental counterpart and therefore $\mathcal{S}^\star(f)=\mathcal{S}(f)$.
For the \ac{SCTS} autocorrelation spectrum $\tilde{c}^{\Vert}_{K}$ the model spectrum $\mathcal{S}^{\Vert}$ is defined analogously but, as explained below, generally does not exactly correspond to an experimentally accessible quantity.
\begin{figure}[t!]
    \includegraphics{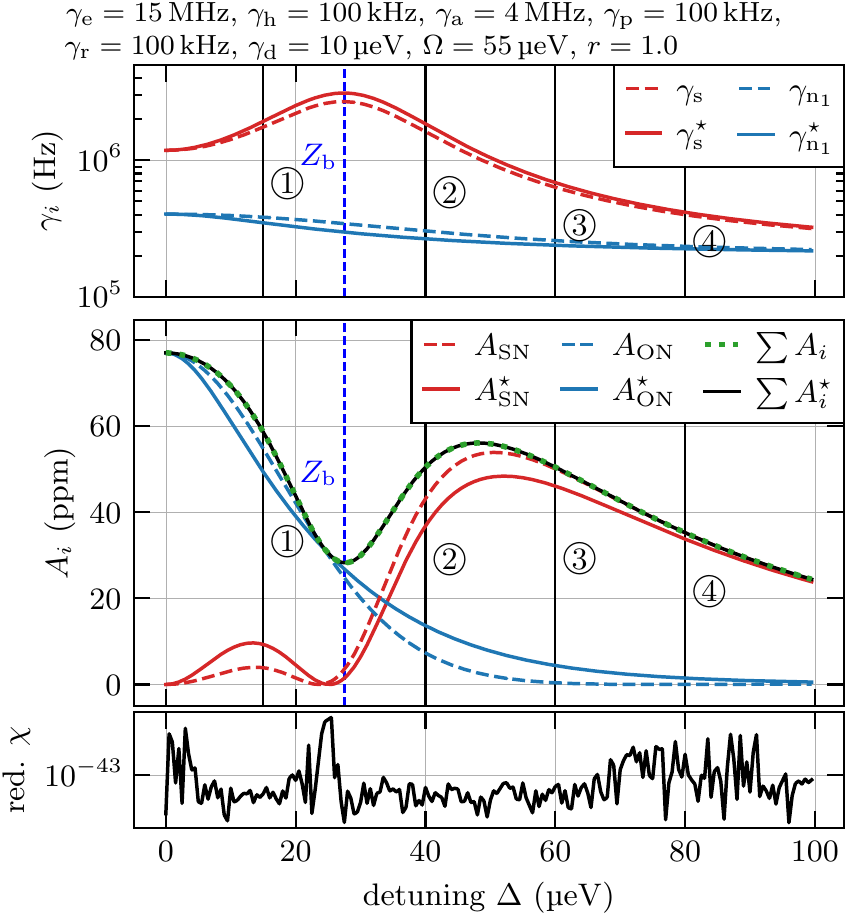}
    \caption{\label{fig:exact_failure}
        For the indicated model parameters, the upper two panels display the spectrum parameters from the \ac{SCTS} (dashed lines: \textcolor{red}{$\gammaS $}, \textcolor{blue}{$\gammaNp $}, \textcolor{red}{$A_{\mathrm{SN}}$}, \textcolor{blue}{$A_{\mathrm{ON}}$}) and their corresponding double Lorentzian regression parameters for a regression of the exact spectrum solution (solid lines: \textcolor{red}{$\gammaS ^{\star}$}, \textcolor{blue}{$\gammaNp ^{\star}$}, \textcolor{red}{$A_{\mathrm{SN}}^{\star}$}, \textcolor{blue}{$A_{\mathrm{ON}}^{\star}$}).
        The regression parameters can significantly deviate while the regression remains excellent (bottom panel).
        The fluctuation of $\chi$ is purely due to numerical noise.
        Nevertheless, the total powers \textcolor{green}{$\sum A_{i}$} and \textcolor{black}{$\sum A_{i}^{\star}$} stay the same.}
\end{figure}

\section{Regression Breakdown}\label{sec:regression_breakdown}
The evaluation and interpretation of the experimental Kerr rotation noise spectra is straightforward in the regime where the SCTS approximation is valid.
Here the separation ansatz allows extracting the distinct spectrally integrated spin and occupation noise powers and the respective correlation rates (i.e., $A_\mathrm{SN}^{\star}, A_\mathrm{ON}^{\star}, \gammaS^{\star}$, and $\gammaNp^{\star}$) from each individual noise spectrum.
We denote this detuning-dependent set of parameters in the following as $P_L^\star(\Delta)$.
On the contrary, the exact solution can only be used to extract the underlying experimental quantities $P_M=\left(\gammaE, \gammaH, \gammaA, \dots\right)$ through a global regression of a set of spectra.
In the following, we study the conditions under which the separation ansatz is a good approximation and where it breaks down.

We have shown in \secref{sec:model_and_theory} that the analytical structures of $\tilde{c}^{\Vert}_{K}$ and $\tilde{c}_{K}$ are quite different.
However, the double Lorentzian spectrum of the \ac{SCTS} solution approximates the shape of the exact solution surprisingly well.
Figure~\ref{fig:exact_failure_example} depicts the calculated spectra according to the exact model (blue lines) and according to the \ac{SCTS} approximation (red line) for three different detunings.
The comparison of the exact model and the approximation shows nearly perfect agreement for $\Delta = 15~\mu$eV at lower frequencies and $\Delta = 60~\mu$eV while the spectra deviate significantly for $\Delta = 40~\mu$eV, i.e., the approximation is in this case excellent at small and large detunings but fails at intermediate detunings.
In order to test if this breakdown of the approximation is directly observable from a single experimental noise spectrum, we also performed fits to the exact solution by a double Lorentzian spectrum for each detuning.
The results of these fits are depicted for all three detunings as black dots in \figref{fig:exact_failure_example}.
Naturally, the fits are perfect for $\Delta = 15~\mu$eV and $\Delta = 60~\mu$eV but, surprisingly, the fit is also perfect for $\Delta = 40~\mu$eV.
This has two interlinked consequences.
First, a single experimental noise spectrum does not indicate if the \ac{SCTS} approximation is valid.
Second, the usually deployed double Lorentzian fit function might perfectly reproduce the shape of the measured noise spectra, but the resulting regression parameters, i.e., $P_L^{\star}(\Delta)$, might deviate from the detuning dependent parameter set $P_L(\Delta)$ predicted by the \ac{SCTS} for the underlying set of model parameters $P_M$.

Figure~\ref{fig:exact_failure} shows this observation in more detail.
The solid lines depict the spin and occupation parameters extracted by fitting a double Lorentzian spectrum to each spectrum calculated by the exact model.
The dashed line represents for comparison the corresponding parameters within the \ac{SCTS} approximation.
Both, rates and amplitudes exhibit for most detuning parameters significant discrepancies, while the sums of the integrated noise powers for the distinct approaches are equal for each detuning.
Obviously, the single-spectrum regression fails to reproduce the \ac{SCTS} parameters even though the reduced $\chi$ of the regression remains at a constant low value (see bottom panel of \figref{fig:exact_failure}).

\begin{figure}[t]
    \includegraphics{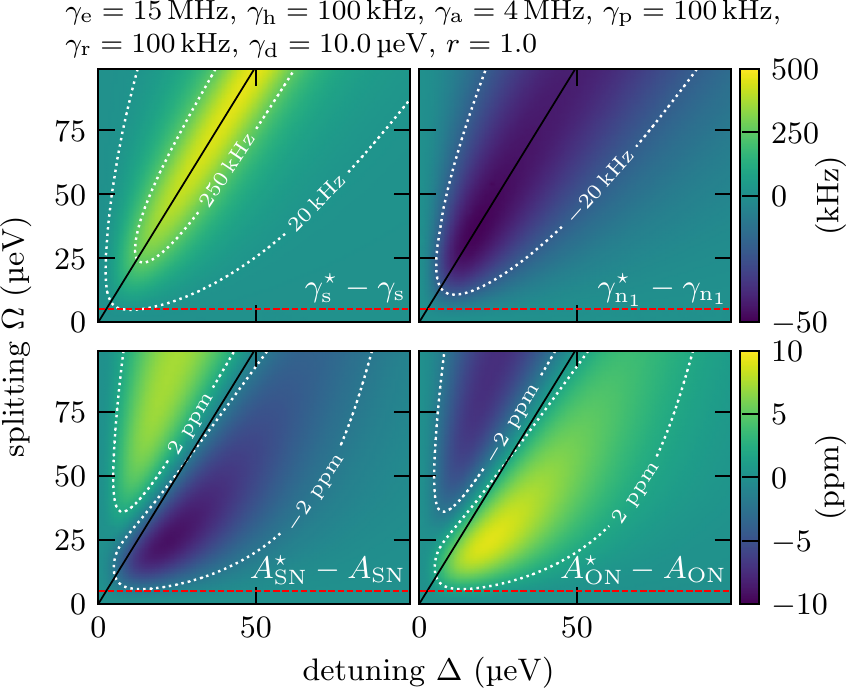}
    \caption{\label{fig:exact_failure_general}Comparison of the regression results and \ac{SCTS} parameters for the specified model parameters.
        Depicted is a smooth interpolation for a grid of $100\times100$ regressions.
        The position of the blue Zeeman branch $Z_{b}$ of the transition is plotted using black diagonal lines.
        White curves indicate selected equipotential lines.
        The color map is shared for each row and is split for the upper panels to consistently display both deviations.
        The \textcolor{red}{red} dashed line indicates the region ($\Omega<\nicefrac{\gammaD }{2}$) below which the \protect\ac{SCTS} remains a good approximation for all detunings.}
\end{figure}
The deviations of the regression parameters are shown in \figref{fig:exact_failure_general} for different detunings and Zeeman splittings.
Here, the differences between the values extracted from a regression to the exact solution and the \ac{SCTS} prediction are plotted for the correlation rates $\gammaS $ and $\gammaNp $ in the upper panels and the noise powers $A_\mathrm{SN}$ and $A_\mathrm{ON}$ in the lower panels.
Only for splittings $\Omega<\nicefrac{\gammaD }{2}$ the \ac{SCTS} solution remains a valid approximation for all detunings (indicated by the red dashed line).
This region corresponds mainly to the range in which previous measurements were made in \citerp{wiegand_spin_2018,sun_non-equilibrium_2022}.
At higher splittings, using a double Lorentzian fit for individual spectra causes the total noise power to be redistributed between the perceived contributions, while the correlation rates are either over- or underestimated.
In the highlighted regions, occupancy and spin dynamics become correlated to a higher degree, and therefore the \ac{SCTS} approximation breaks down.

\begin{figure}[t]
    \includegraphics{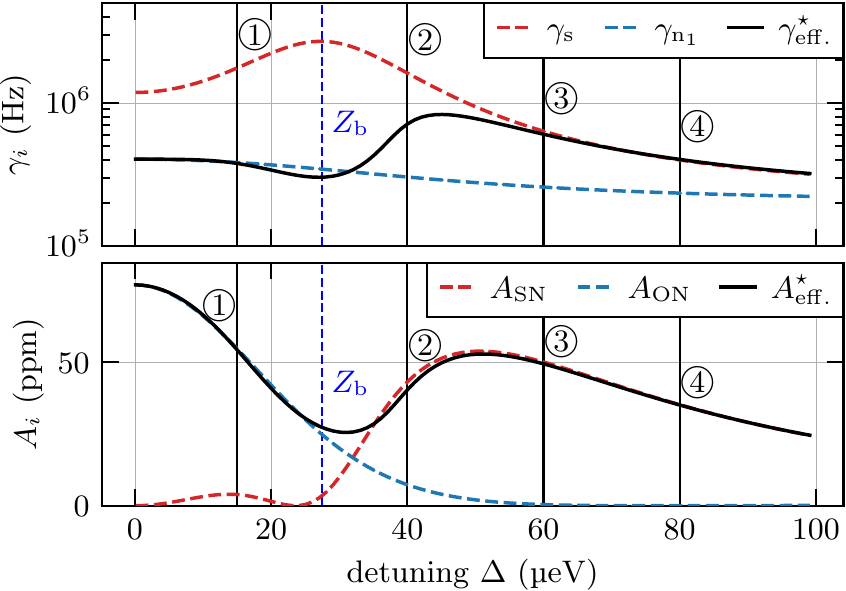}
    \caption{\label{fig:exact_failure_zero}Single Lorentzian regression of the exact solution for the same model parameters as in \figref{fig:exact_failure}.
        Again, the upper two panels display the spectrum parameters from the \protect\ac{SCTS} (dashed lines: \textcolor{red}{$\gammaS $}, \textcolor{blue}{$\gammaNp $}, \textcolor{red}{$A_{\mathrm{SN}}$}, \textcolor{blue}{$A_{\mathrm{ON}}$}) and the corresponding single Lorentzian regression parameters (solid black: $\gammaEff ^{\star}$, $A_\mathrm{eff.}^{\star}$).
        The regression parameters converge towards one of the contributions for $\Delta\approx0$ and $\bigl|\Delta\bigr|\gg\nicefrac{\Omega}{2}$.}
\end{figure}
However, the \ac{SCTS} solution remains valid in two regions where only a single contribution dominates (see \figref{fig:exact_failure_zero}).
Immediately around $\Delta\approx0$ only an ON contribution is present, whereas for large detunings $\Delta\gg\nicefrac{\Omega}{2}$ only an SN contribution remains.
A \emph{single} Lorentzian fit that disregards the other contribution yields a much better approximation of the relevant \ac{SCTS} parameters.
Therefore, these limiting cases can be used to extract useful information from \emph{individual} spectra for any set of model parameters:
\begin{align}
    \lim_{\Delta\rightarrow0}\gammaEff ^{\star}      & = \lim_{\Delta\rightarrow0}\gammaNp ^{\star}
    = \lim_{\Delta\rightarrow0}\gammaNp \nonumber                                                                                                                                \\
                                                     & =\gammaR +\gammaX +\frac{r}{\left(r+1\right)}\frac{2\gammaD ^{2}}{4\gammaD ^{2}+\Omega^{2}}\left(\gammaA -\gammaX \right).
    \label{eq:DMF-asymp-zero}
    \\
    \lim_{\Delta\rightarrow\infty}\gammaEff ^{\star} & = \lim_{\Delta\rightarrow\infty}\gammaS ^{\star}
    = \lim_{\Delta\rightarrow\infty}\gammaS \nonumber                                                                                                                                 \\
                                                     & =\gammaH +\gammaX
    \label{eq:DMF-asymp-inf}.
\end{align}

To conclude this section, we note that a similar regression breakdown can occur for experimental spectra even when the \ac{SCTS} solution works in an ideal case.
Any minute deformation in the analyzer's spectral response will lead to the same kind of redistribution between contributions - this is a general inherent weakness of the single-spectrum regression compared to global fitting methods.

\section{Model Parameterization}\label{sec:parameterization}
\begin{figure*}[t]
    \includegraphics{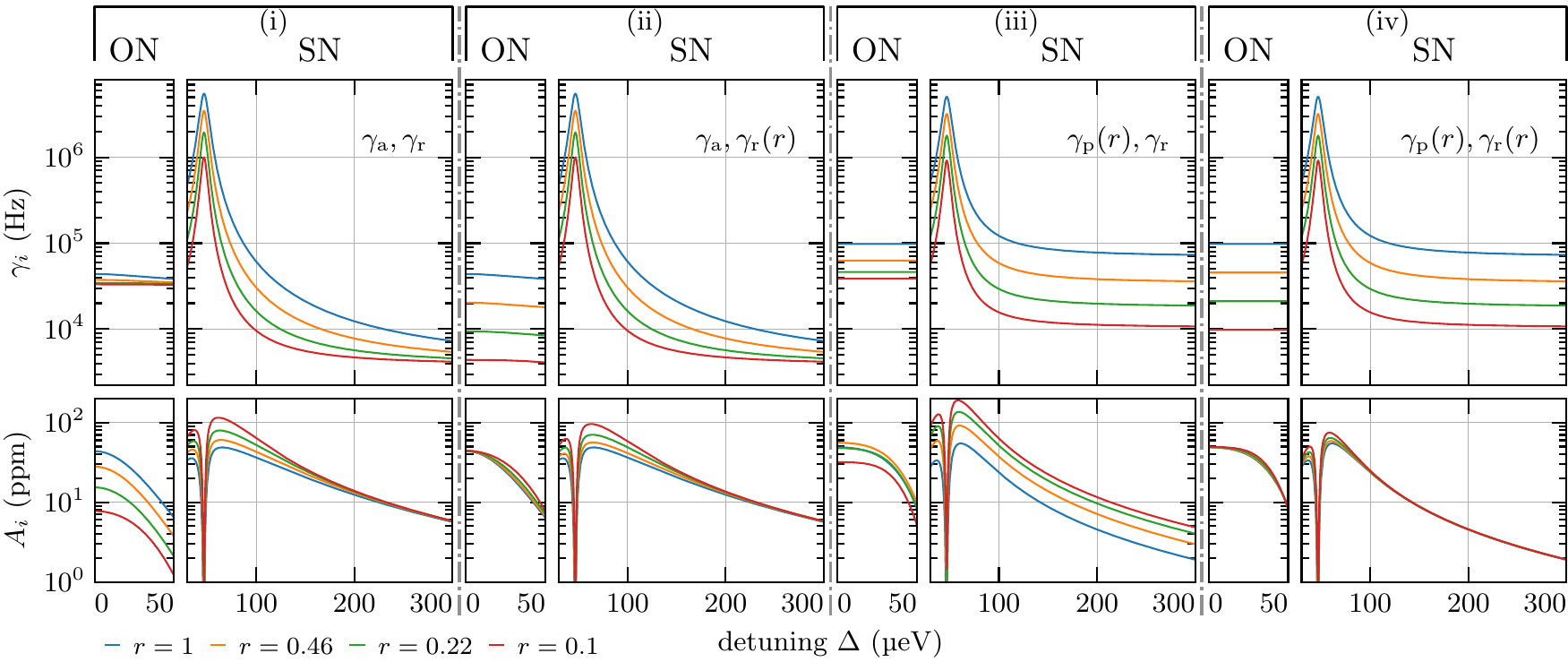}
    \caption{\label{fig:exact_failure_tmp}Comparison of different model parameterizations.
        Rates without an explicitly stated dependence are assumed to be fixed (i.e, $\gammaA $).
        Rates with a specified $r$ dependence (i.e., $\gammaR (r)$) are assumed to be proportional to $r$ without any offset.
        Only the scenario with photoeffect-like depopulation and repopulation (rightmost column) describes the experimental data in \figref{fig:correlation_rates_and_noise_powers} and \figref{fig:model_and_data} correctly.}
\end{figure*}
Although the predicted experimental parameters obtained from the \ac{SCTS} solution are not always perfect estimates for a real spectrum, they can be used qualitatively to gain insight into the probe intensity dependence of the underlying model parameters and thereby identify the relevant physical processes of the QD dynamics.
To this end, we consider different possible scenarios for the hole dynamics, which are summarized in \figref{fig:exact_failure_tmp}.
The figure is divided into four columns (i) to (iv), and each column represents different assumptions for the magnitude or the probe intensity dependence of the model parameters $\gammaA $, $\gammaR $, and $\gammaX $.
We will compare the depicted detuning dependences to our measurement results in the subsequent experimental section.
Each column in \figref{fig:exact_failure_tmp} consists of four panels: the left and right panels display the \ac{SCTS} parameters for the correlation rate and noise power of the respective ON and SN contributions.

For the first column (i), the \ac{QD} hole occupancy is depleted with an Auger rate rate $\gammaA $ from the excited states, while the \ac{QD} ground states are replenished through a constant rate of reoccupation $\gammaR $, e.g., by tunneling from a nearby acceptor state.
This is the case that has been assumed in \citerp{wiegand_spin_2018,wiegand_hole-capture_2018}.
Characteristic for this case is that a) the spin correlation rate $\gammaS $ converges towards $\gammaH $ for high detunings regardless of the probe intensity, b) the correlation rate $\gammaNp $ is peak-shaped around $\Delta = 0$, and c) the ON power $A_\mathrm{ON}$ has a pronounced probe laser intensity dependence.

For the second column (ii), the constant reoccupation rate $\gammaR $ is replaced by an intensity-dependent reoccupation rate $\gammaR (r)\propto r$, where $r$ is again the relative intensity.
Equation~(\ref{eq:DMF-asymp-zero}) directly shows that such a linear intensity dependence of the $\gammaR $ rate yields a linear intensity dependence of the asymptotic value of $\gammaNp $.
As a second consequence, for $\Delta \approx 0, r > 0$ the intensity dependence
of the occupancy noise power is almost canceled out, and hence the noise power remains at a more or less constant value.

For the third column (iii), the Auger rate is set to zero ($\gammaA =0$) and instead a probe intensity dependent hole loss from the \emph{ground} states is introduced, $\gammaX (r)\propto r$, while the reoccupation rate $\gammaR $ is again assumed to be constant. The charge loss from the excited state can be neglected as the trion decay rate $\gammaTr$ is fast and carrier tunneling from the trion state is suppressed \citep{kurzmann_optical_2016}.
We can see that the spin correlation rate $\gammaS $ converges in this case to an intensity-dependent value given by \eqref{eq:DMF-asymp-zero}, while $\gammaNp $ is a flat line with an intensity-dependent offset that has almost no detuning dependence.
The noise powers of both contributions increase in this scenario with the probe intensity.

Finally, for the fourth column (iv), \emph{both} $\gammaX (r)\propto r$ and $\gammaR (r)\propto r$ are assumed to be linearly intensity dependent, whereas no charge loss from excited states is present ($\gammaA =0$).
Here, like in the second column, a variable $\gammaR (r)$ causes the ON power to be preserved across different probe intensities, but now this is in good approximation also the case for the SN power.
Again, the correlation rates $\gammaS $ and $\gammaNp $ exhibit a clear linear dependence on the probe laser intensity.

\section{Experiment}\label{sec:experiment}
\begin{figure}[t!]
    \includegraphics{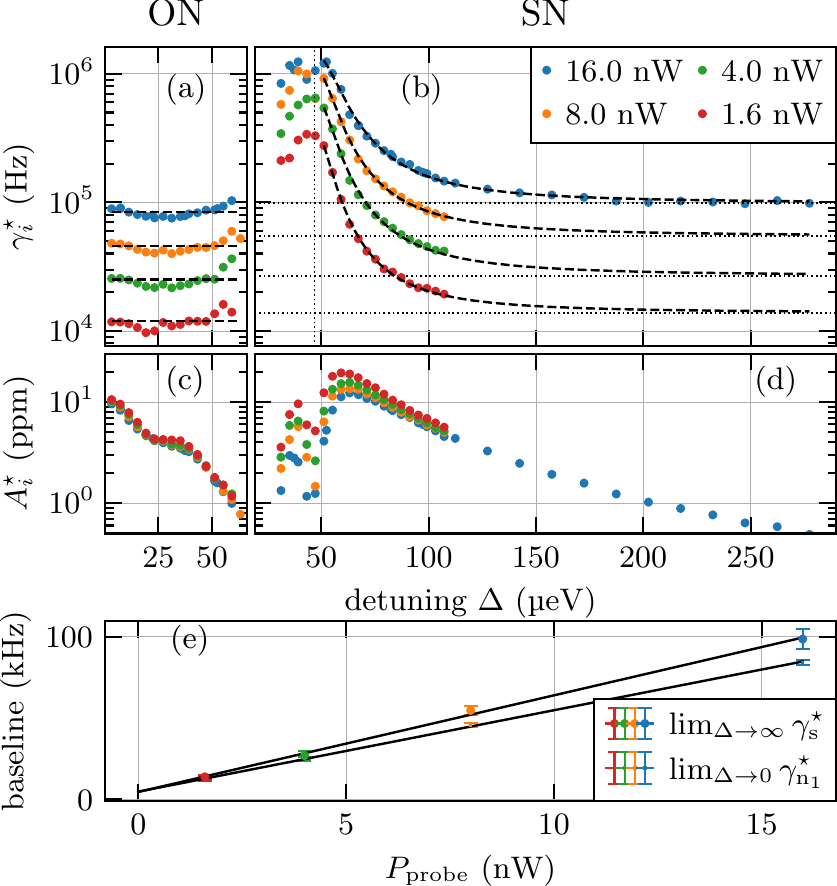}
    \caption{\label{fig:correlation_rates_and_noise_powers}%
        Asymptotic rates at various probe powers, $\SI{1.8}{\kelvin}$ lattice temperature, and $\SI{750}{\milli\tesla}$ external magnetic field.
        Panels a) and c) display the extracted parameters for $\gammaNp ^{\star}$ and $A_\mathrm{ON}^{\star}$ of the narrow ON contribution.
        Panels b) and d) show the respective parameters $\gammaS ^{\star}$ and $A_\mathrm{SN}^{\star}$ for the broad SN contribution.
        Both correlations rates display a visible linear laser power dependence of the asymptotically approached range as summarized in panel e).
        The dashed lines in panel a) designate the the mean of $\gammaNp ^{\star}$ For panel b) single Lorentzian curves on top of a constant baseline (dotted line) with their center fixed at the central position of the Zeeman branch are fitted.}
\end{figure}

In the following, we measure Kerr rotation noise spectra of a single hole localized in a single (InGa)As QD and compare the experimental results with the four scenarios depicted in \figref{fig:exact_failure_tmp} in order to identify the relevant physical processes for the QD depopulation and repopulation dynamics.
We use the same sample as in \citerp{wiegand_hole-capture_2018,wiegand_spin_2018,dahbashi_measurement_2012,dahbashi_optical_2014} which contains a single layer of low density (InGa)As \acp{QD} embedded in a low-finesse, asymmetric Bragg cavity with 13 and 30 AlAs/GaAs $\nicefrac{\lambda}{4}$ pairs as top and bottom mirrors, respectively.
The bottom mirror allows us to probe the sample in reflection geometry, while the top mirror increases the effective number of round trips of the probe laser and thereby the Kerr noise signal \citep{kavokin_resonant_1997}.
The structure is grown by molecular beam epitaxy with a residual background p-type doping of $\approx$ $\SI{1e14}{\cm^{-3}}$.
This doping results in a finite probability of the \acp{QD} being occupied at thermal equilibrium by a single hole.
The QD density is well below $\SI{1}{\mu m^{-2}}$ such that individual \acp{QD} can be easily addressed with our confocal microscope, which focuses the laser light down to a beam diameter of $\approx\SI{1}{\mu m}$.
Further details concerning the sample structure, the identification of charged QDs, and the spin noise spectroscopy setup are described in \citerp{wiegand_hole-capture_2018,wiegand_spin_2018,dahbashi_measurement_2012,dahbashi_optical_2014}.
Only one important experimental detail has been changed.
Usually, spin-noise background spectra have been measured by changing the magnetic field.
Such an approach is tedious for high magnitude fields produced by superconducting coils since significant changes of the magnetic field become prohibitively slow.
Therefore, background spectra have been measured by moving the relevant QD horizontally out of the laser focus by a fast, high-precision, low-temperature piezo scanner.

All measurement results presented in this publication are recorded for an individual \ac{QD} during a single cool-down.
This particular \ac{QD} is selected for its rather large splitting of $\approx\SI{758}{\micro\electronvolt}$ at $B=\SI{0}{\tesla}$ between the trion ($\mathrm{X^{+}}$) and neutral ($\mathrm{X}$) exciton resonances.
The large splitting allows one to easily separate any noise contributions of the neutral exciton.
The spin noise measurements presented in the following are recorded at a magnetic field of $\SI{750}{\milli\tesla}$, a resulting Zeeman splitting of $\SI{93+-1}{\micro\electronvolt}$, and a lattice temperature of $\SI{1.8}{\kelvin}$ which corresponds to a thermal energy of $\approx\SI{155}{\micro\electronvolt}$.
The splittings of the individual hole and electron (pseudo-) spin systems are smaller than the total splitting and we therefore assume that the spin temperature is high enough to disregard any polarization effects.

In the following, we present Kerr rotation noise measurements in dependence on detuning for four different laser intensities.
The noise spectra are evaluated within the \ac{SCTS} approximation, and the resulting intensity and detuning dependences are compared in respect to the four scenarios depicted in \figref{fig:exact_failure_tmp} and discussed in \secref{sec:parameterization}.
The four panels in the top of \figref{fig:correlation_rates_and_noise_powers} show the experimental results in the same format as the individual scenarios in \figref{fig:exact_failure_tmp}.
For the experimental data, each single spectrum has been fitted for $\num{30} \le \Delta \le \SI{63}{\micro\electronvolt}$ by a double Lorentzian fit and outside of this region by a single Lorentzian.
Naturally, the curves presented in \figref{fig:correlation_rates_and_noise_powers} potentially suffer from the aforementioned regression breakdown.
Therefore, we do not consider the SN component in the region $\Delta < \SI{50}{\micro\electronvolt}$ for the further evaluation.
For the remaining values, we perform a nonlinear regression of the values of $\gammaS^\star$ using a Lorentzian, i.e., $\kappa_{+}$, on top of a constant baseline.
The resulting baseline value gives an estimate of $\lim_{\Delta\rightarrow \infty}\gammaS^{\star}$.
The detuning dependence of the ON correlation rate is flat, and we take the mean of the values as an estimate of $\lim_{\Delta\rightarrow 0}\gammaNp^{\star}$.
The results of these secondary regressions (bottom panel of \figref{fig:correlation_rates_and_noise_powers}) may not be extremely accurate, but they clearly reveal that the correlation rates of both, the ON and SN contributions show a linear probe intensity dependence for $\Delta \rightarrow 0$ and $\Delta \rightarrow \infty$, respectively.
Furthermore, the ON power shows nearly no and the SN power only a very weak intensity dependence.
This behavior corresponds to the prediction by the fourth column of \figref{fig:exact_failure_tmp}, i.e, both deoccupation and reoccupation of the QD by a hole are determined by a laser driven photoeffect.
\begin{figure}[t!]
    \includegraphics{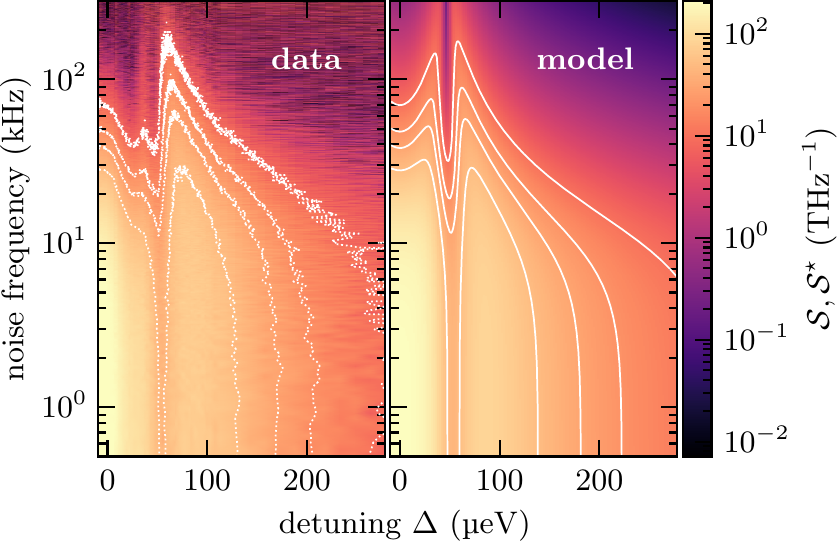}
    \caption{\label{fig:model_and_data}%
        Comparison of model and raw data for a data set recorded at $\SI{1.8}{\kelvin}$ lattice temperature, $\SI{16}{\nano\watt}$ probe power, and $\SI{750}{\milli\tesla}$ external longitudinal magnetic field. The \ac{PSD} scale is normalized by the square of the probe power. Only positive detunings are shown and the data was clipped in the same range as the model.}
\end{figure}

Next, we compare the data measured at the highest intensity with the exact solution.
We assumed in \secref{sec:model_and_theory} a rather simplistic (exponentially decaying) outer state dynamics.
This assumption gives a qualitatively correct spectral shape for a region, where the magnitude of occupancy fluctuations is small.
However, the ON is the main contribution in the region between the Zeeman branches and, in this case, even the ``exact'' solution cannot describe the spectrum perfectly as depicted in \figref{fig:model_and_data} for $\Delta < \SI{50}{\micro\electronvolt}$.
Consequently, quantitative statements concerning $\gammaE $ and $r$ are associated with a large error.
Since we have no means in the current setup to determine these values from complementary measurements, we use as realistic estimations for the calculations in \figref{fig:model_and_data} a relative intensity of $r=1$ for the highest probe intensity of $\SI{16}{\nano\watt}$ and an electron relaxation rate of $\gammaE =\SI{40}{\mega\hertz}$.
The hole relaxation rate $\gammaH $ is determined as $\gammaH \approx\SI{3.83+-0.15}{\kilo\hertz}$ by performing an extrapolation towards zero probe intensity at a fixed detuning $\Delta > \SI{100}{\micro\electronvolt}$.
The remaining parameters can be extracted from the experiment, though their accuracy depends on the accuracy of the assumed parameters.
A global regression (for $\SI{0}{\micro\electronvolt} \le\Delta \le \SI{10}{\micro\electronvolt}$ and $\Delta \ge \SI{50}{\micro\electronvolt}$) of the respective experimental spectra yields:
$\gammaD  \approx \SI{3.6188 +- 0.0022}{\micro\electronvolt}$,
$\gammaR  \approx \SI{32.00 +- 0.07}{\kilo\hertz}\times r$ and
$\gammaX  \approx \SI{66.13 +- 0.07}{\kilo\hertz}\times r$, where the uncertainties are given by the regression error and do not represent the accuracy errors that are in the order of $\gtrsim{\SI{1}{\micro\electronvolt}}$ and $\gtrsim{\SI{1}{\kilo\hertz}}$, respectively.

\section{Conclusion}\label{sec:conclusion}
We have extended the theoretical framework of Kerr rotation noise (spin noise spectroscopy) of single semiconductor quantum dots to high magnetic fields.
At low magnetic fields, the separation of correlator time scales is a good approximation and simplifies the extraction of spin and charge relaxation rates and powers.
At high magnetic fields, on the other hand, the approximation fails for certain regimes of parameters and the dynamics becomes more complex since the noise spectrum cannot be separated anymore in spin and charge noise, i.e., the evaluation and interpretation of the experimental data requires some caution.
What is more, the comparison between theory and experiment on unbiased (InGa)As QDs suggests that even the full theoretical model does not fully describe the occupation noise dynamics at low detunings because of the rather simplistic assumptions of the outer-state dynamics.
Cause is the intricate charging dynamics by a finite number of acceptor states in the QD surrounding which is different from QD to QD.
This variation is probably the most important reason why the properties of unbiased (InGa)As QDs with respect to quantum photonic devices vary strongly even for closely spaced QDs on the same waver.
Nevertheless, the presented theory allows to extract reliable conclusions concerning the underlying physical process of charging and recharging and concerning the spin relaxation dynamics from the experimental data.
Most importantly, the intensity and detuning dependences of the experimental data show clearly that both, decharging and recharging of the (InGa)As QD by a hole are dominated by incoherent, two-way, laser-driven photoeffect-like processes.
This again is important for quantum photonic devices since p-type background doping by carbon impurities is unavoidable in molecular beam epitaxy of GaAs.
As a consequence, all unbiased (InGa)As QD photonic devices will be ultimately limited by this two-way photoeffect-like charging effect.

Previous measurements at low magnetic field were strongly influenced by the Auger effect which could be suppressed by large laser detunings in order to measure the prevailing moderately long hole spin relation times.
The current experiment at higher magnetic fields indicates that the Auger effect itself is significantly reduced by high magnetic fields and that the hole spin relaxation becomes extremely long.
Quantitative measurements of such extremely long spin relaxation times are challenging since the photoeffect-induced incoherent charge dynamics opens an effective pseudospin relaxation channel even at large detunings.
In fact, the extremely long hole spin relaxation times suggest future spin noise measurements on a single-photon level.
Combination of such single-photon spin noise measurements with resonance fluorescence measurements would definitely increase the accuracy of \emph{quantitative} estimates of parameters and might even allow one to give more profound insight into the individual two-way charging dynamics of single QDs.

\section{Acknowledgements}
We thank K. Pierz (PTB) for providing the sample and M. M. Glazov (Ioffe Institute) for fruitful discussions. This work was funded by the Deutsche Forschungsgemeinschaft (DFG, German Research Foundation) under Germany’s Excellence Strategy – EXC-2123 QuantumFrontiers – 390837967 and OE 177/10-2.

P.S. and K.H. contributed equally to this work.

\bibliographystyle{apsrev4-1}
\bibliography{references}

\end{document}